\documentclass[conference]{IEEEtran}
\IEEEoverridecommandlockouts
\usepackage{cite}
\usepackage{amsmath,amssymb,amsfonts}
\usepackage{algorithmic}
\usepackage{graphicx}
\usepackage{textcomp}
\usepackage{xcolor}
\usepackage{subcaption}
\usepackage[numbers]{natbib}
\def\BibTeX{{\rm B\kern-.05em{\sc i\kern-.025em b}\kern-.08em
    T\kern-.1667em\lower.7ex\hbox{E}\kern-.125emX}}
\begin{document}

\title{Job-level Carbon and Water Footprint Estimation for HPC: Bias Assessment from Runtime to Full Life Cycle
}

\author{\IEEEauthorblockN{Xi Chen}
\IEEEauthorblockA{\textit{LIACS, Leiden University} \\
Leiden, Netherlands \\
x.chen@liacs.leidenuniv.nl}
\and
\IEEEauthorblockN{Chris Broekema}
\IEEEauthorblockA{\textit{ASTRON} \\
Dwingeloo, Netherlands \\
broekema@astron.nl}
\and
\IEEEauthorblockN{Rob van Nieuwpoort}
\IEEEauthorblockA{\textit{LIACS, Leiden University} \\
Leiden, Netherlands \\
r.v.van.nieuwpoort@liacs.leidenuniv.nl}

}

\maketitle

\begin{abstract}
High performance computing evaluation has traditionally focused on performance and energy, but these metrics alone cannot capture the sustainability cost of runtime configurations. We proposes a unified job-level water and carbon accounting framework with both operational and embodied impacts. Results show that higher thread counts generally reduce total footprint, but the benefit diminishes at higher thread counts. Water is mainly dominated by embodied impact, whereas carbon is mainly dominated by operational impact.
\end{abstract}

\begin{IEEEkeywords}
high performance computing, water footprint, carbon footprint, embodied impact, runtime configuration
\end{IEEEkeywords}

\section{Introduction}
HPC (High Performance Computing) benchmarks have long focused on performance, throughput, and energy use as key metrics. However, as sustainability becomes more important, using only energy is no longer enough to fully assess the environmental cost of individual (HPC) jobs on different system configurations.

Existing studies show that sustainability discussions for HPC and data centers have moved beyond simple energy efficiency. They now include carbon emissions, water use, and some life cycle impacts \citep{holbling2025energy, lei2025water, chouksey2026green}. However, prior work typically examines carbon, water, and broader life-cycle impacts separately, rather than assigning them together to individual HPC jobs within a single accounting boundary. As a result, even when the same benchmark runs on the same hardware, the sustainability impact of different runtime configurations cannot be directly compared under a unified job-level water–carbon view.

To address this, we propose a job-level water–carbon accounting framework for HPC workloads. Our approach combines empirical measurements of runtime and energy consumption with model-based accounting of water and carbon impacts. Under a single analysis scope, it jointly characterizes operational water, embodied water, operational carbon, and embodied carbon, and maps them to specific runtime configurations.




\section{Methodology}
This section presents the proposed job-level accounting framework for jointly estimating water and carbon impacts of HPC runs. 
For a given run, total impacts are defined as:
\begin{equation}
W_{\mathrm{total}} = W_{\mathrm{op}} + W_{\mathrm{emb}}, \quad
C_{\mathrm{total}} = C_{\mathrm{op}} + C_{\mathrm{emb}}.
\end{equation}
Here, $W_{\mathrm{op}}$ and $C_{\mathrm{op}}$ denote operational water and carbon, while $W_{\mathrm{emb}}$ and $C_{\mathrm{emb}}$ denote embodied impacts from hardware manufacturing and packaging~\citep{jiang2025thirstyflops, li2023toward}. 

Operational water is modeled as the sum of direct cooling water and indirect water from electricity generation:
\begin{equation}
W_{\mathrm{op}} = E \cdot \mathrm{WUE}(T_{wb}) + E \cdot \mathrm{PUE} \cdot \mathrm{EWF}.
\end{equation}
Here, $E$ is job energy (kWh), $\mathrm{WUE}(T_{wb})$ is water usage effectiveness as a function of wet-bulb temperature, and $\mathrm{EWF}$ is the energy water factor derived from grid energy mix.

Operational carbon is defined as:
\begin{equation}
C_{\mathrm{op}} = E \cdot \mathrm{PUE} \cdot CI_{\mathrm{grid}},
\end{equation}
where $CI_{\mathrm{grid}}$ is the grid carbon intensity. Water and carbon thus share the same energy basis but differ in their environmental intensity factors.

Embodied impacts are amortized at the whole-node level using hardware lifetime and long-term utilization. 

First, total embodied water and carbon are estimated at node level using component-based models from prior work \citep{jiang2025thirstyflops, li2023toward, falk2025more}. Then, they are converted into node-hour rates:
\begin{equation}
r^{W}_{\mathrm{node}}=\frac{W_{\mathrm{emb,node}}}{u_{\mathrm{active}}H}, \quad
r^{C}_{\mathrm{node}}=\frac{C_{\mathrm{emb,node}}}{u_{\mathrm{active}}H},
\end{equation}
where $H$ is hardware lifetime (hours) and $u_{\mathrm{active}}$ is long-term utilization.

Job-level embodied impacts are then:
\begin{equation}
W_{\mathrm{emb}} = r^{W}_{\mathrm{node}} \cdot t_{\mathrm{job}}, \quad
C_{\mathrm{emb}} = r^{C}_{\mathrm{node}} \cdot t_{\mathrm{job}}.
\end{equation}



\section{Experiment}
\subsection{Experimental Setup}
Experiments were conducted on DAS-6 (ASTRON). The system is a dual-socket Intel Xeon Gold 6140 CPU platform (2.30\,GHz) with 18 physical cores per socket, 2 hardware threads per core, and 2 NUMA nodes, for a total of 36 physical cores and 72 logical CPUs.

OpenMP thread counts were set to \(\{1,2,4,8,16,18,36\}\), covering low-thread execution, intermediate scaling points, full utilization of one socket (18 threads), and full utilization of both sockets with physical cores only (36 threads).

Three benchmarks were evaluated: STREAM Triad, a memory-bandwidth-oriented benchmark with performance reported in MB/s, and the compute-oriented OpenMP NAS Parallel Benchmarks CG.C and EP.C, with performance reported in Mop/s. All benchmark runs were executed through a common wrapper around ASTRON's \texttt{pmt} measurement tool, which was used to record execution time, total energy, and average power; the wrapper itself only standardized benchmark invocation and log collection. The resulting CSV logs were then extended with sustainability metrics, including operational, embodied, and total carbon footprint, as well as direct, indirect, operational, embodied, and total water footprint.

Each thread setting was repeated multiple times to reduce noise: 3 runs for STREAM Triad and 2 runs each for CG.C and EP.C. This resulted in 49 runs in total.

\subsection{Results}
\begin{figure}[htbp]
\centering
\begin{subfigure}{0.48\linewidth}
    \centering
    \includegraphics[width=\linewidth]{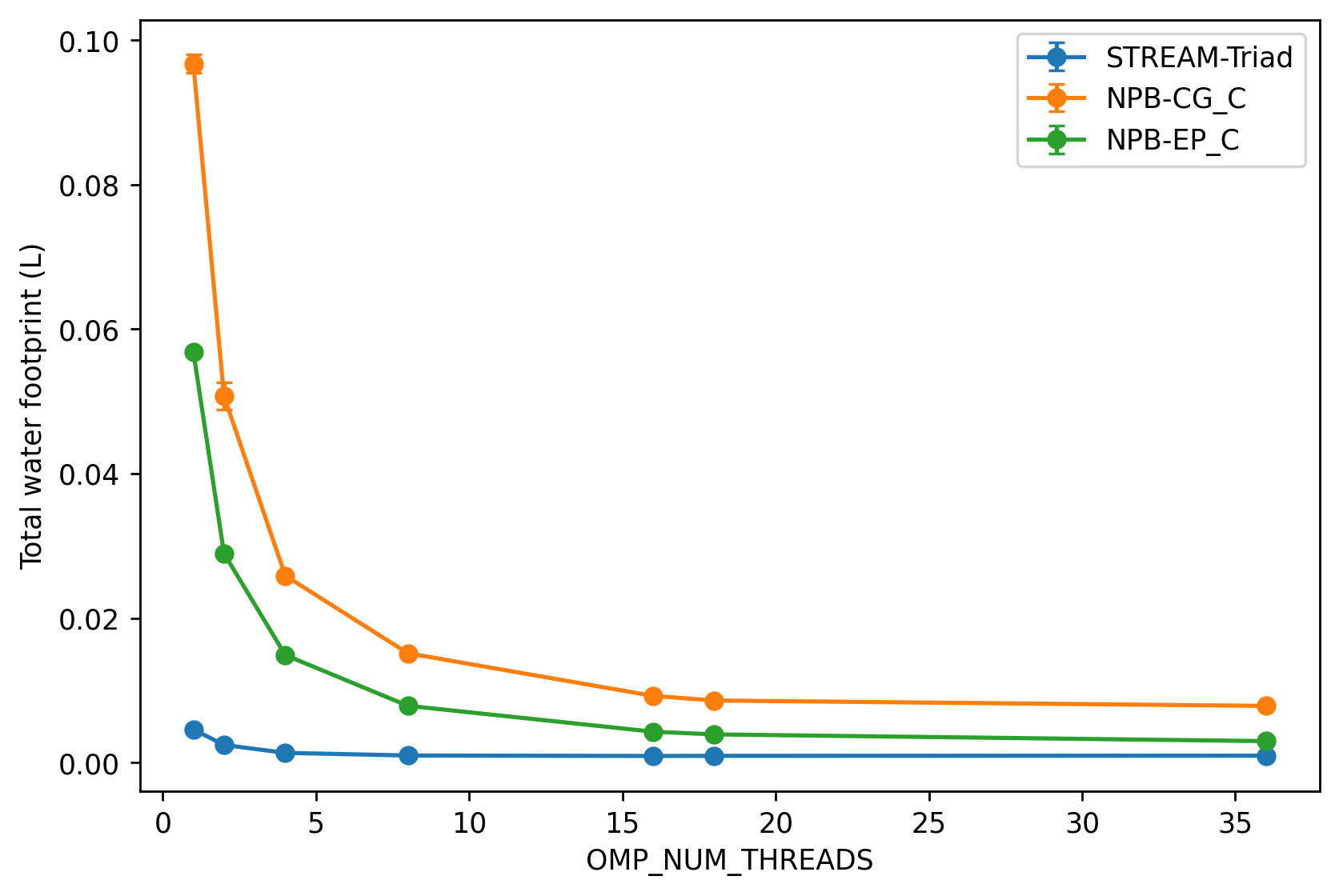}
    \caption{Water}
    \label{fig:totalWater}
\end{subfigure}
\hfill
\begin{subfigure}{0.48\linewidth}
    \centering
    \includegraphics[width=\linewidth]{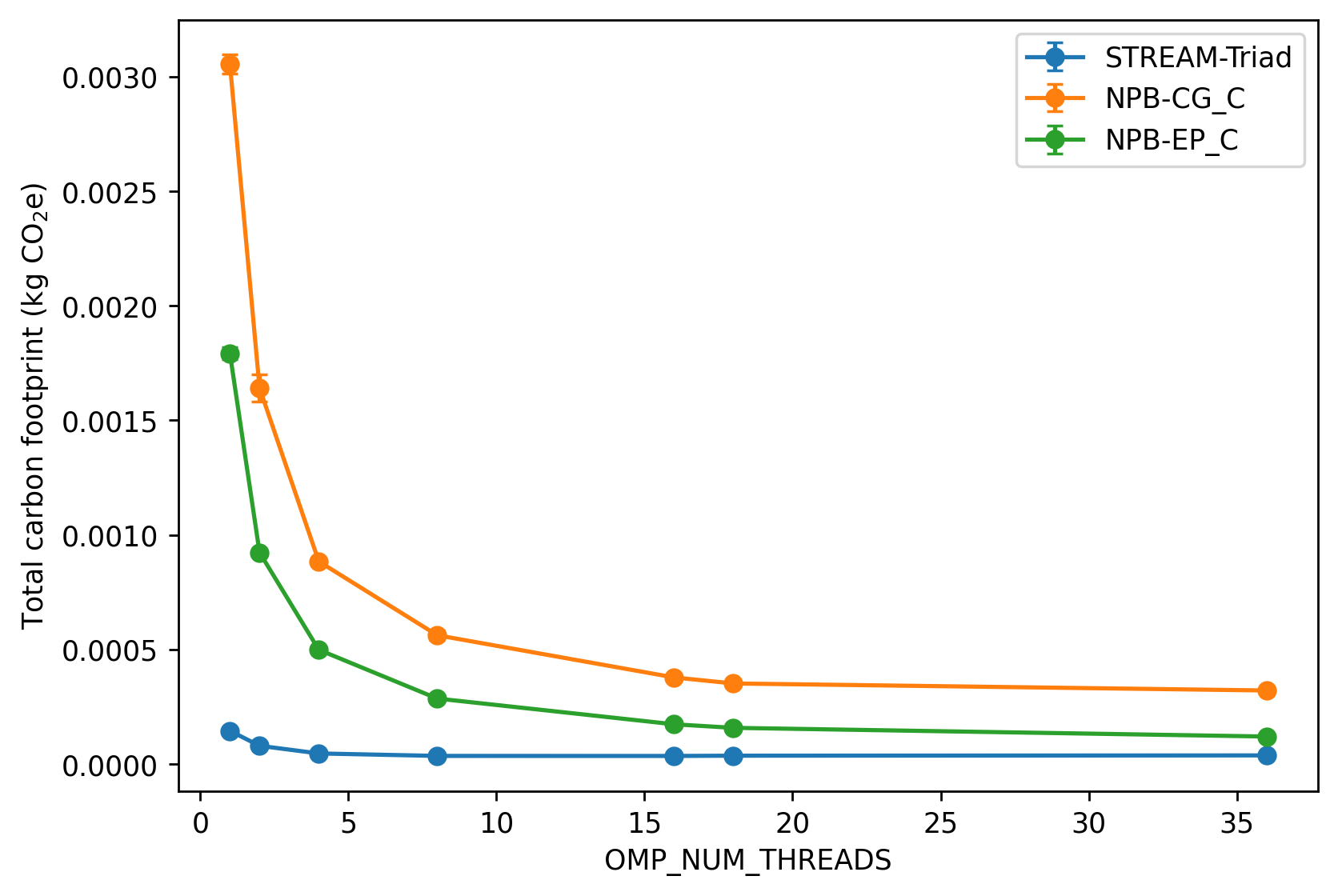}
    \caption{Carbon}
    \label{fig:totalCarbon}
\end{subfigure}
\caption{Total water footprint (left) and total carbon footprint (right) versus OpenMP thread count for STREAM-Triad, NPB-CG\_C, and NPB-EP\_C. For all three benchmarks, increasing the thread count reduces total footprint, but the improvement becomes marginal at higher thread counts.}
\label{fig:total_wc_threads}
\end{figure}

Fig.~\ref{fig:total_wc_threads} shows that increasing the thread count generally lowers both total water and total carbon footprint across all three benchmarks because the run times decrease (the run-to-idle effect), although the benefit diminishes at higher thread counts.

\begin{figure}[htbp]
\centering
\includegraphics[width=\linewidth]{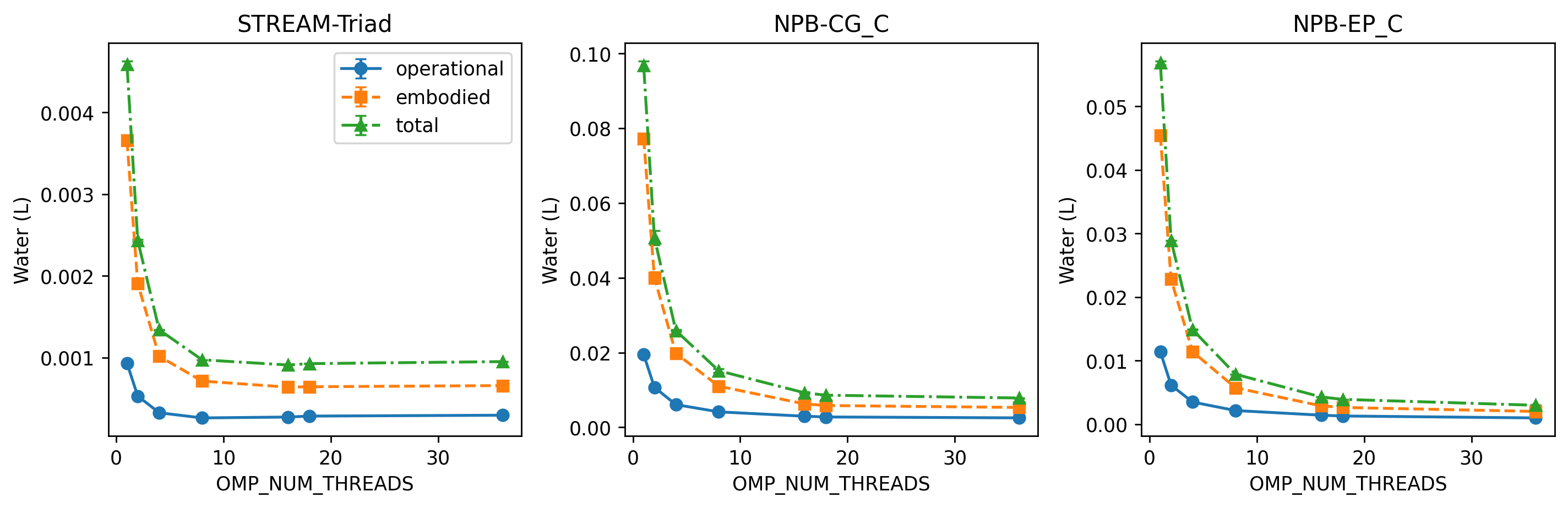}
\caption{Operational, embodied, and total water footprint versus OpenMP thread count. Embodied water remains higher than operational water across all benchmarks and all tested configurations, indicating that operational-only accounting would systematically underestimate total water impacts.}
\label{fig:water_breakdown_threads}
\end{figure}

\begin{figure}[htbp]
\centering
\includegraphics[width=\linewidth]{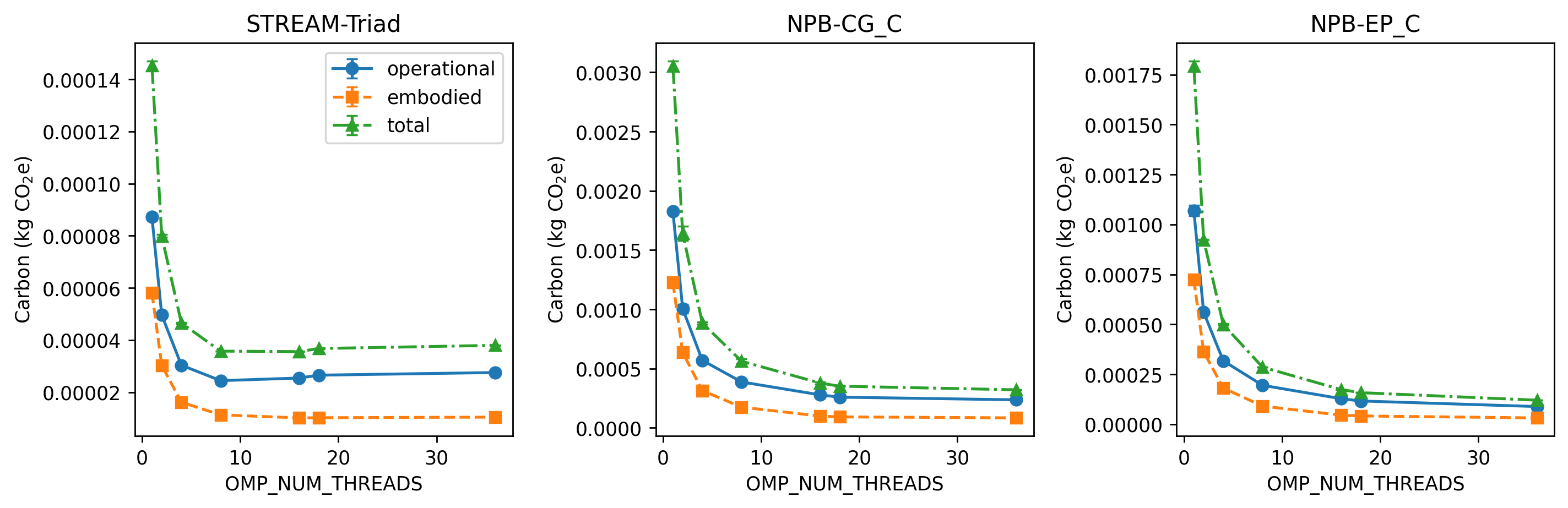}
\caption{Operational, embodied, and total carbon footprint versus OpenMP thread count. In contrast to water, operational carbon remains the dominant component across all benchmarks, while embodied carbon adds a smaller but non-negligible contribution.}
\label{fig:carbon_breakdown_threads}
\end{figure}

Figs.~\ref{fig:water_breakdown_threads} and \ref{fig:carbon_breakdown_threads} show that the balance between operational and embodied impacts is very different for water and carbon. For water, embodied impact is the main component across all tested configurations. In contrast, carbon is still mainly dominated by the operational phase. In both cases, increasing the thread count reduces the total footprint, mainly because the runtime becomes shorter. However, the relative contribution of different lifecycle stages remains different.

This contrast is important for configuration-level assessment. An operational-only analysis would greatly underestimate total water impact, and would also underestimate total carbon impact, but to a smaller extent. Overall, thread scaling improves sustainability mainly in the low-to-mid thread range. The improvement becomes limited when performance approaches saturation.


\section{Conclusion}
This paper presents a job-level accounting framework to jointly estimate water and carbon footprint in HPC benchmark runs. By integrating operational and embodied impacts in a unified configuration-level pipeline, the framework allows comparison of runtime settings under a consistent sustainability scope, instead of using only performance and energy.

Experiments on DAS-6 with STREAM-Triad, NPB-CG\_C, and NPB-EP\_C show that increasing thread count generally reduces both total water and total carbon footprint. However, the benefit becomes small at higher thread counts as scaling approaches saturation. The results also show a clear difference between impact types. In this setup, water footprint is mainly dominated by embodied impact, while carbon footprint is still mainly dominated by the operational phase. As a result, simplified accounting choices can bias the interpretation at the configuration level, especially when water impacts are estimated using operational-only methods.

Overall, these findings suggest that sustainability-aware HPC benchmarking should not treat accounting scope as a secondary reporting detail. Whether operational and embodied impacts are included, and how they are assigned to jobs, can directly affect how runtime configurations are interpreted and compared. Future work will extend the analysis to more compute-intensive CPU workloads and study how shared-node execution and different allocation rules affect job-level sustainability accounting.

\section*{Acknowledgment}
This work was funded by the OTP research project "SuperCode: SUstainability PER AI-driven CO-DEsign", project number 2025/TTW/01878134, which is financed by the Dutch Research Council (NWO) domain Applied and Engineering Sciences (TTW).

\bibliographystyle{IEEEtran}
\bibliography{references}

\end{document}